\documentclass{PoS}

\def\aap{A\&A}
\def\aj{AJ}%
\def\araa{ARA\&A}%
\def\apj{ApJ}%
\def\apjl{ApJ}%
\def\apjs{ApJS}%
\def\aap{A\&A}%
\def\aapr{A\&A~Rev.}%
\def\mnras{MNRAS}%
\def\nar{New A Rev.}%
\def\pasa{PASA}%
\title{New constraints on binary evolution enhance the supernova type Ia rate}

\ShortTitle{Supernova Type Ia rates}

\author{\speaker{S. Toonen}\\
        Leiden Observatory, Leiden University, PO Box 9513 Leiden, The
Netherlands\\
Anton Pannekoek Institute for Astronomy, University of Amsterdam, NL-1090 GE Amsterdam, the Netherlands\\
        E-mail: \email{toonen@uva.nl}}

\abstract{Even though Type Ia supernovae  (SNIa) play an important role in many fields in astronomy, the nature of the progenitors of SNIa remain a mystery. One of the classical evolutionary pathways towards a SNIa explosion is the single degenerate (SD) channel, in which a carbon-oxygen white dwarf accretes matter from its non-degenerate companion until it reaches the Chandrasekhar mass. Constraints on the contribution from the SD channel to the overall SNIa rate come from a variety of methods, e.g. from  abundances, from signatures of the companion star in the light curve or near the SNIa remnant, and from synthetic SNIa rates. 
In this proceedings, I show that when incorporating our newest understandings of binary evolution, the SNIa rate from the single degenerate channel is enhanced. I also discuss the applicability of these constraints on the evolution of SNIa progenitors.
}

\FullConference{The Golden Age of Cataclysmic Variables and Related Objects - III, Golden2015 \\
		7-12 September 2015\\
		Palermo, Italy}

\begin{document}

\section{Introduction}
\label{sec:intro}
Despite the pivotal role that type Ia supernovae (SNIa) have played in our understanding of the universe and its expansion, a theoretical explanation of the origin of these events is still lacking. There is a consensus that SNIa are thermonuclear explosions of white dwarfs, however it is unclear how the explosion can be triggered (see \cite{Wan12} for a review). One of the classical channels is the single degenerate (SD) channel \cite{Whe73}. In this scenario a non-degenerate star transfers hydrogen-rich material to a white dwarf, which steadily increases in mass until the Chandrasekhar mass is reached. 
In reality, efficient mass growth of the white dwarf may be limited to a rather narrow range in mass transfer rates around $10^{-7}\,\rm M_{\odot} \rm yr^{-1}$. At these rates hydrogen is burned continuously on the surface of the white dwarf \cite{Nom82}.
At lower mass transfer rates, the accreted material is burned is a thermonuclear runaway  \cite{Sch50, Sta74}. 
In these nova eruptions, some or all of the previously accreted matter is ejected from the system, and possibly even some of the surface material of the white dwarf is ejected (e.g. \cite{Pri95}). At high mass accretion rates close to the Eddington limit, the growth of the white dwarf is limited. At these rates a giant-like envelope develops around the white dwarf that can interact with the donor star or from which strong winds can develop \cite{Kat94, Hac96}. The uncertainties in the efficiency of mass accretion onto white dwarfs gives rise to a significant systematic error in the predicted SNIa rates of the SD channel \cite{Bou13}.

\section{New constraints on binary evolution}
During the formation of  many compact binaries, the systems evolve through one or more common-envelope (CE) phases (e.g. \cite{Too14}). 
During this phase, both stars are engulfed in a single envelope. If enough energy and angular momentum is available to drive of the envelope, a merger of the two stars can be avoided \cite{Pac76}. What remains, is a binary with one or two compact objects in a tight orbit. 
Despite the dramatic effect of CE-phase on binary evolution, it is not understood in detail (see e.g. \cite{Iva13} for a review). Only few constraints on CE-evolution exist, which rely on observed samples of close binaries and theoretical modelling of binary evolution. 

The most recent constraint on CE-evolution is based on observations of post-common envelope binaries (PCEBs). These are detached binaries with a white dwarf and a main-sequence component in a compact orbit. 
The periods of these binaries have been observed to range from a few hours to a few days \cite{Neb11}, however, theory has predicted periods up to hundreds of days. 
By tracing the evolution of the observed PCEBs (from the SDSS) backwards in time, 
Zorotovic et al. \cite{Zor10} has shown that the CE-phase leads to a stronger shrinkage of the orbit than previously assumed. With a different method, Toonen et al. \cite{Too13} shows that the surplus of synthetic PCEBs at long periods can not be explained by observational biases; Even though the sample of PCEBs is severely affected by selection effects against finding a dim white dwarf next to a relatively  bright main-sequence star, the period distribution of PCEBs is not affected significantly by observational biases. The authors conclude that in order to reconcile the theoretical and observed period distribution, the orbital separation has to decrease significantly during the CE-phase. With similar methods, the studies of \cite{Cam14} and \cite{Zor14} confirm the results of \cite{Too13}. 

In quantitative terms, the CE-phase is often described by conservation of energy \cite{Tut79, Web84}
\begin{equation}
E_{\rm gr} = \alpha \Delta E_{\rm orb},
\label{eq:alpha-ce}
\end{equation}
where $E_{\rm gr}$ is the binding energy of the envelope, $\Delta E_{\rm orb}$ the difference in orbital energy before and after the CE-phase, and $\alpha$ describes the efficiency with which orbital energy can be used to unbind the CE. The binding energy of the envelope is estimated by:
\begin{equation}
E_{\rm gr} = \frac{GMM_{\rm env}}{\lambda R},
\label{eq:egr}
\end{equation}
 where $M$ is the mass of the star, $M_{\rm env}$ is the mass of the envelope, $R$ is the radius of the star, and $\lambda$ is the structure parameter of the star.
Conservative use of energy has been commonly assumed in binary evolution calculations ($\alpha=1$), however, a value of $\alpha\sim 0.25$ is more appropriate for PCEBs (as found by the studies described above).

\section{Effect on SNIa rate}

Using the binary population synthesis code SeBa \cite{Por96, Too12}, I model the evolution of a large number of binaries from the zero-age main-sequence  including processes such as stellar winds and mass transfer. The models presented here are based on a Kroupa IMF \cite{Kro93} for the primary masses, a uniform mass ratio distribution, a log-uniform distribution in orbital separations \cite{Abt83}, a thermal eccentricity distribution \cite{Heg75} and a binary fraction of 50\%. I adopt two models for the CE-phase that differ in the CE-efficiency; $\alpha\lambda=2$ in model $\alpha\alpha$ and $\alpha\lambda = 0.25$ in model $\alpha\alpha 2$. 
See \cite{Too13} for more details on the models.

\begin{figure}[h!]
\caption{\textbf{Delay-time distribution} SNIa rate as a function of time assuming a single burst of star formation at $t=0$ for model $\alpha\alpha$ (blue solid line) and model $\alpha\alpha2$ (green dashed line). The SNIa rate is increased for the model with a lower efficiency of orbital energy usage during the common-envelope phase. }
\centering
\includegraphics[width=0.8\columnwidth]{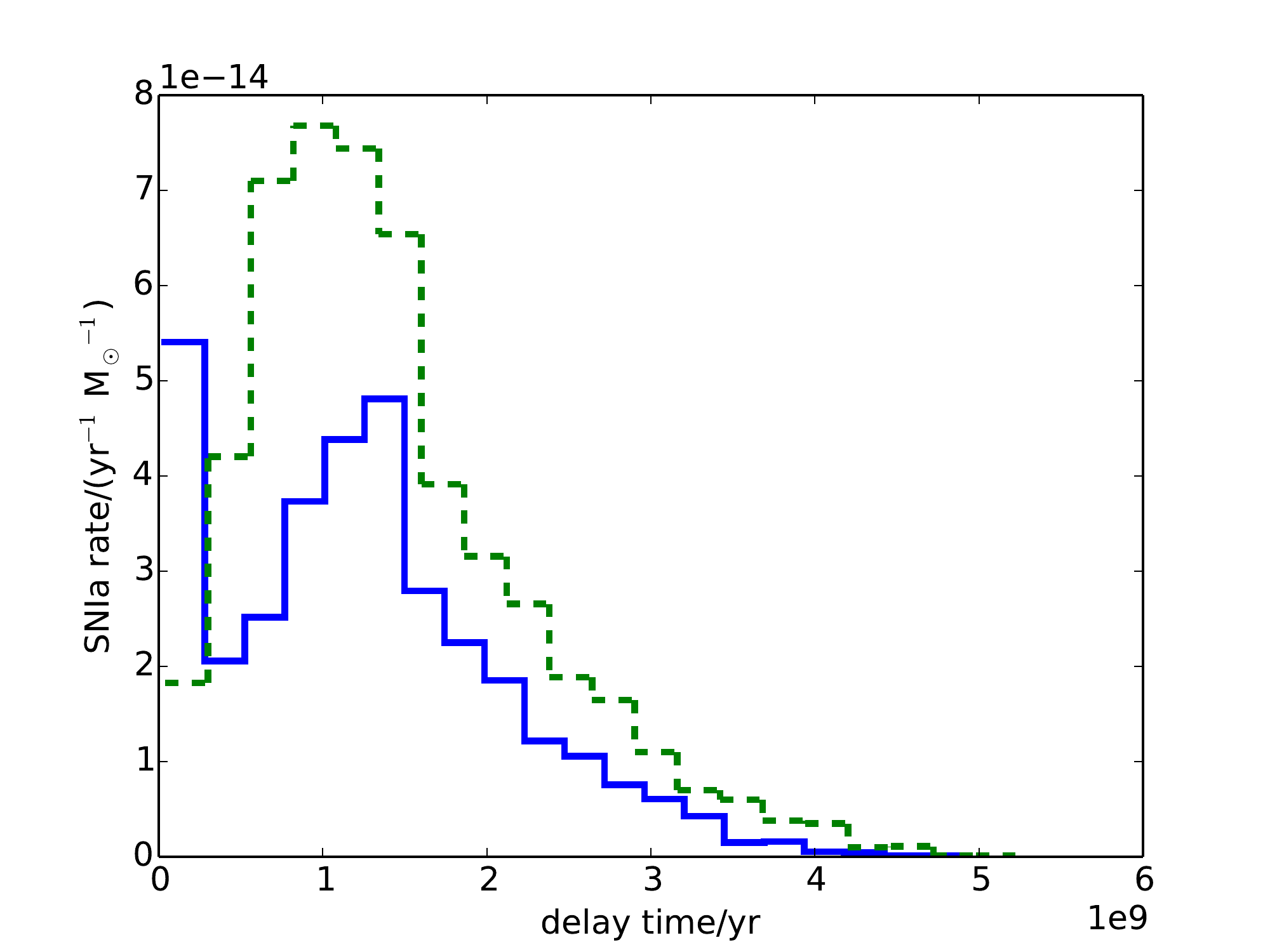} 
\label{fig:dtd}
\end{figure}

I find that the SNIa rate is enhanced in model $\alpha\alpha2$ compared to model $\alpha\alpha$ i.e. when the CE-efficiency is low. The total time-integrated number of SNe Ia per unit formed stellar mass is $1.34\cdot 10^{-4}\,\rm M_{\odot}^{-1}$ and $0.835\cdot 10^{-4}\,\rm M_{\odot}^{-1}$. The SNIa rate as a function of time  (Fig.\ref{fig:dtd}) show similar characteristics for both models. The SD SNIa rate peak at short delay times, and show a sharp decline towards longer delay times. No SNIa explosions are expected from the classical SD channel after several Gyr (but see \cite{DiS11, Jus11}).

The enhancement is in qualitative agreement with the results of  Claeys et al.\cite{Cla14} and Ruiter et al. \cite{Rui09}. In the standard model of \cite{Cla14} ($\alpha =1$, variable $\lambda$) the integrated SNIa rate is $0.087\cdot 10^{-4}\,\rm M_{\odot}^{-1}$. This is an order of magnitude below our predictions, which is likely due to the sensitivity of the accretion efficiency to the mass transfer rate (see section \ref{sec:intro} and \cite{Bou13}). When decreasing the CE-efficiency to $\alpha=0.5$ and $\alpha=0.2$, \cite{Cla14} find an increase in the SNIa rate of a factor 1.5 respectively a factor 10(!). Ruiter et al. \cite{Rui09} finds that the integrated SNIa rate increases by about a factor 3 if the CE-efficiency is decreased from $\alpha\lambda =1$ to $\alpha\lambda=0.5$. In a subsequent paper \cite{Rui11}, these authors also study a model with $\alpha\lambda = 0.125$, but unfortunately they do not specify the time-integrated rate. For comparison, the observed SNIa rate is of the order of $10^{-3}\,\rm M_{\odot}^{-1}$ \cite{Mao12,Gra13,Per12}.

The reason for the increase in the SNIa rate due to the decreased CE-efficiency is elegant. For a lower CE-efficiency, the progenitors of SD SNIa arise from wider binaries. With standard assumptions for the initial distributions of binary parameters (as described above), fewer systems are born at larger orbital separations. 
So naively, one would expect fewer SNIa explosions if the CE-efficiency is low.
However, the evolution of the SD SNIa progenitors are not the same in model $\alpha\alpha$ and model $\alpha\alpha2$. If a SNIa progenitor evolves from a wider system, the stars have more time to grow their cores before the onset of the CE-phase. Thus on average the white dwarfs will be more massive. This is helpful because 1) the white dwarf has to accrete less material to reach the Chandrasekhar mass 2) mass accretion onto a more massive white dwarf is more efficient (e.g. \cite{Hac08,Bou13}) 3) the mass transfer phase with the white dwarf as accretor is more likely to be dynamically stable. As a result, the average mass of the donor star is higher for model $\alpha\alpha2$ compared to model $\alpha\alpha$, see Fig.\ref{fig:island}.

\section{Conclusion and Discussion}
In this letter, the supernova type Ia rate is studied from binaries in which a white dwarf accretes matter from a non-degenerate companion star (single degenerate channel). Studies of post-common envelope binaries have shown that common-envelope evolution leads to a stronger shrinkage of the period than what has been assumed previously. If this is also the case for the SNIa progenitors in the single degenerate 
channel, the SNIa rate is increased.

\begin{figure}
    \centering
    \begin{tabular}{c c}
	\includegraphics[scale = 0.45]{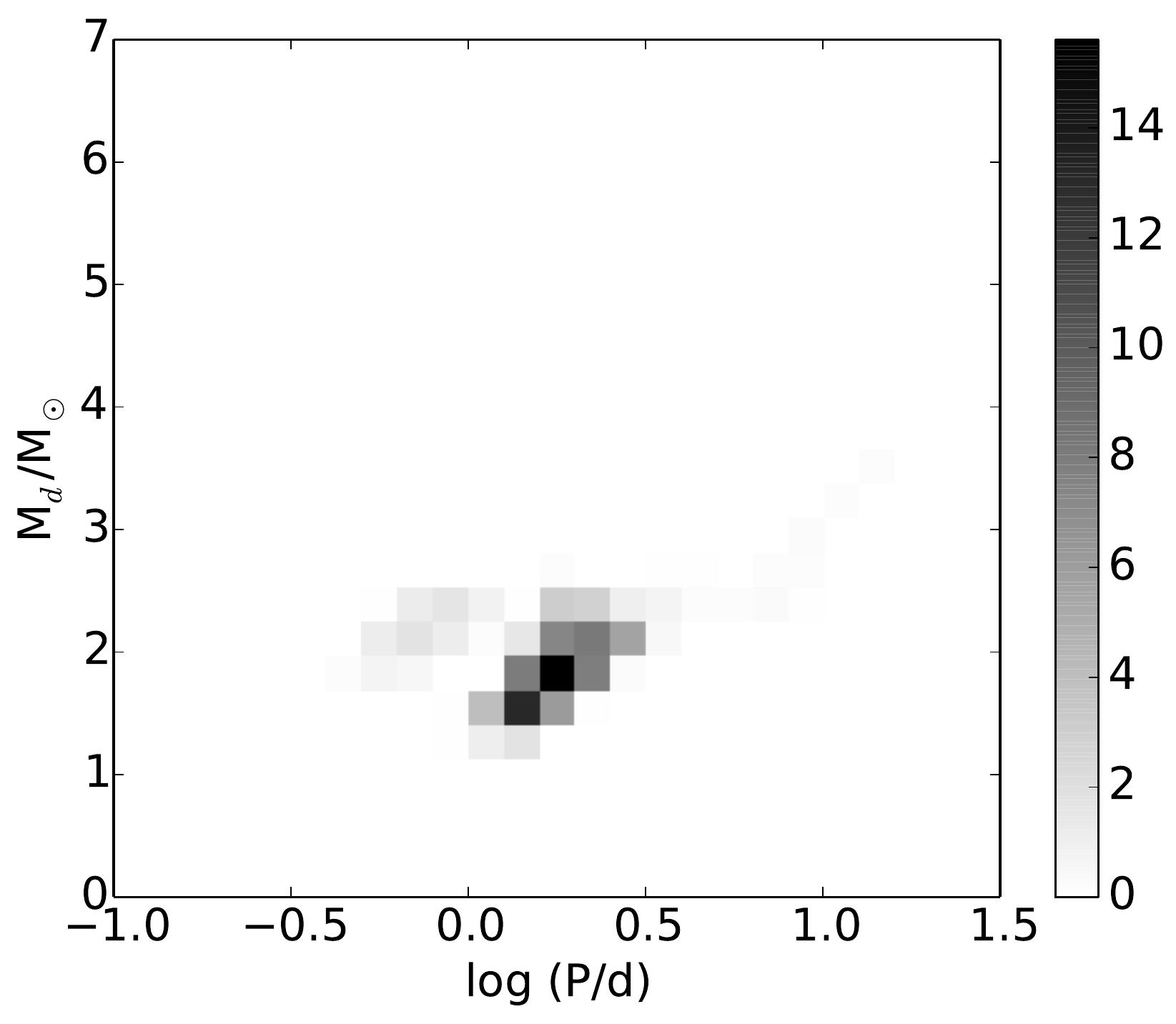} &
	\includegraphics[scale = 0.45]{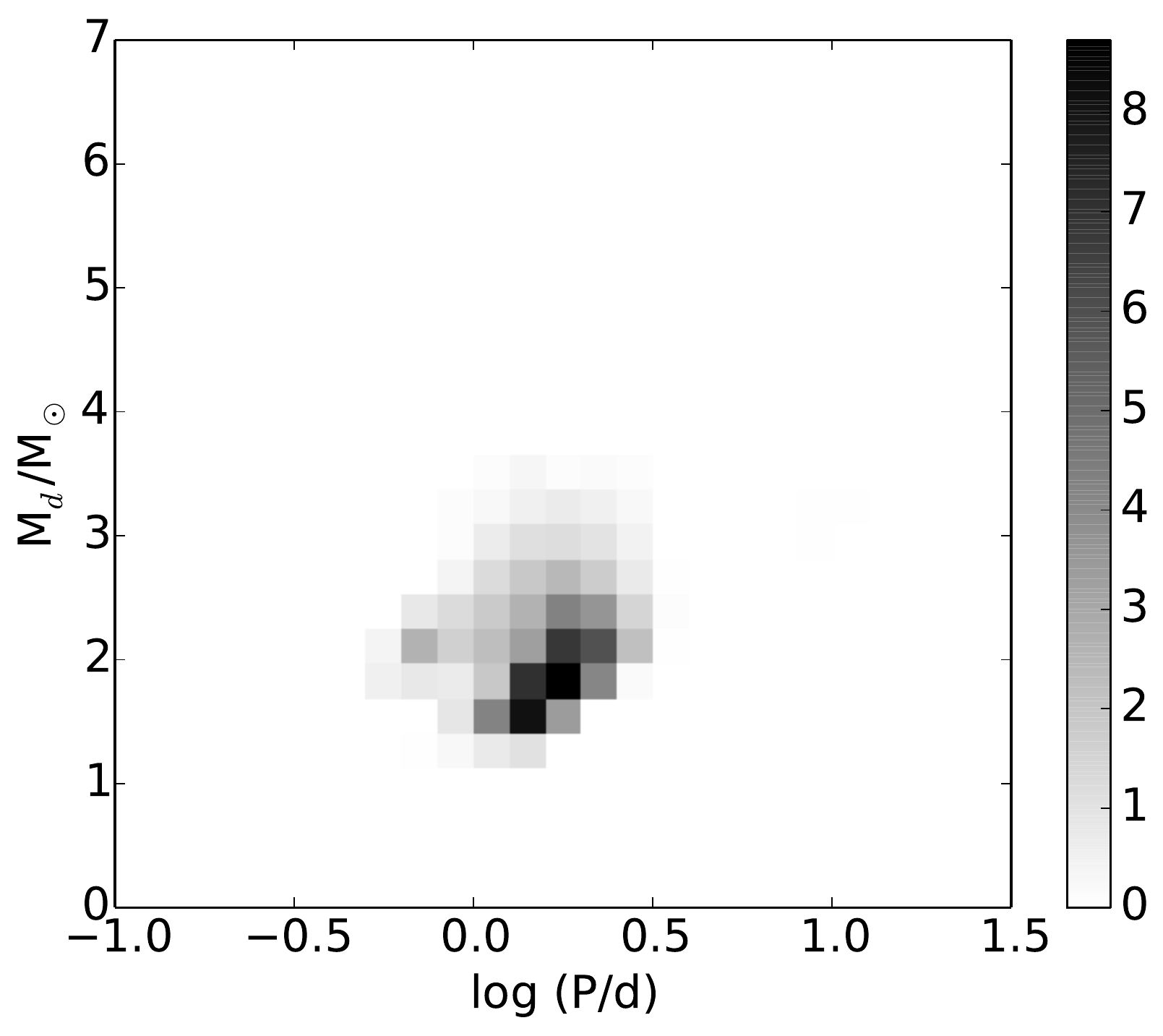} \\
	\end{tabular}
    \caption{\textbf{Islands} Donor mass vs. orbital period if the SD SNIa progenitors after the last mass-transfer event in which the WD is formed. On the left model $\alpha\alpha$ is shown, on the right model $\alpha\alpha2$. The intensity of the grey scale corresponds to the density of objects on a linear scale in percentages of all SD SNIa progenitors of the corresponding model.  } 
    \label{fig:island}
    \end{figure}

A priori, it is not obvious that the constraint on CE-evolution from post-common envelope binaries is appropriate to SD SNIa progenitors. The mass ratios of the PCEB progenitors are similar to those of the SD SNIa progenitors ($q\sim 0.2-0.5$ and $q\sim 0.2-0.7$, respectively), however the stellar masses involved in the CE-event are different. For example, the masses of the non-degenerate companion of the SD SNIa progenitors are given in Fig.\ref{fig:island}, while for PCEBs the masses are less than $0.5\,\rm M_{\odot}$. Currently, it is unclear how the stellar masses and mass ratios affect the progression of the CE-phase. Further constraints on the CE-phase - either from stellar populations or hydrodynamic simulations - are instrumental in understanding the progenitors of supernova Type Ia, as well as other stellar populations such as X-ray, gamma-ray and gravitational wave sources.

\section{Acknowledgements}
I thank Thomas Wijnen for the helpful discussions. 
This work was supported by the Netherlands Research Council NWO (grant VENI [nr.
639.041.645])


\begin{thebibliography}{99}
\bibitem{Abt83} Abt, H.~A.\ 1983, \araa, 21, 343 
\bibitem{Bou13} Bours, M.~C.~P., Toonen, S., \& Nelemans, G.\ 2013, \aap, 552, A24 
\bibitem{Cam14} Camacho, J., Torres, S., Garc{\'{\i}}a-Berro, E., et al.\ 2014, \aap, 566, A86 
\bibitem{Cla14} Claeys, J.~S.~W., Pols, O.~R., Izzard, R.~G., Vink, J., \& Verbunt, F.~W.~M.\ 2014, \aap, 563, A83 
\bibitem{DiS11} Di Stefano, R., Voss, R., \& Claeys, J.~S.~W.\ 2011, \apjl, 738, L1 
\bibitem{Gra13} Graur, O., \& Maoz, D.\ 2013, \mnras, 430, 1746 
\bibitem{Hac96} Hachisu, I., Kato, M., \& Nomoto, K.\ 1996, \apjl, 470, L97 
\bibitem{Hac08} Hachisu, I., Kato, M., \& Nomoto, K.\ 2008, \apj, 679, 1390-1404 
\bibitem{Heg75} Heggie, D.~C.\ 1975, \mnras, 173, 729 
\bibitem{Iva13} Ivanova, N., Justham, S., Chen, X., et al.\ 2013, \aapr, 21, 59 
\bibitem{Jus11} Justham, S.\ 2011, \apjl, 730, L34 
\bibitem{Kat94} Kato, M., \& Hachisu, I.\ 1994, \apj, 437, 802 
\bibitem{Kro93} Kroupa, P., Tout, C.~A., \& Gilmore, G.\ 1993, \mnras, 262, 545 
\bibitem{Mao12} Maoz, D., \& Mannucci, F.\ 2012, \pasa, 29, 447 
\bibitem{Neb11} Nebot G{\'o}mez-Mor{\'a}n, A., G{\"a}nsicke, B.~T., Schreiber, M.~R., et al.\ 2011, \aap, 536, A43 
\bibitem{Nom82} Nomoto, K.\ 1982, \apj, 253, 798 
\bibitem{Pac76} Paczynski, B.\ 1976, Structure and Evolution of Close Binary Systems, 73, 75 
\bibitem{Per12} Perrett, K., Sullivan, M., Conley, A., et al.\ 2012, \aj, 144, 59 
\bibitem{Por96} Portegies Zwart, S.~F., \& Verbunt, F.\ 1996, \aap, 309, 179 
\bibitem{Pri95} Prialnik, D., \& Kovetz, A.\ 1995, \apj, 445, 789 
\bibitem{Rui09} Ruiter, A.~J., Belczynski, K., \& Fryer, C.\ 2009, \apj, 699, 2026 
\bibitem{Rui11} Ruiter, A.~J., Belczynski, K., Sim, S.~A., et al.\ 2011, \mnras, 417, 408 
\bibitem{Sch50} Schatzman, E.\ 1950, Annales d'Astrophysique, 13, 384 
\bibitem{Sta74} Starrfield, S., Sparks, W.~M., \& Truran, J.~W.\ 1974, \apjs, 28, 247 
\bibitem{Too12} Toonen, S., Nelemans, G., \& Portegies Zwart, S.\ 2012, \aap, 546, A70 
\bibitem{Too13} Toonen, S., \& Nelemans, G.\ 2013, \aap, 557, A87 
\bibitem{Too14} Toonen, S., Claeys, J.~S.~W., Mennekens, N., \& Ruiter, A.~J.\ 2014, \aap, 562, A14 
\bibitem{Tut79}	Tutukov, A., \& Yungelson, L.\, Mass loss and evolution of O-type stars; Proceedings of the
Symposium, Vancouver Island, British Columbia, Canada, June 5-9, 1978.
(A80-16501 04-90) Dordrecht, D. Reidel Publishing Co., 1979, p. 401-
406; Discussion, p. 407

\bibitem{Wan12} Wang, B., \& Han, Z.\ 2012, \nar, 56, 122 
\bibitem{Web84} Webbink, R.~F.\ 1984, \apj, 277, 355 
\bibitem{Whe73} Whelan, J., \& Iben, I., Jr.\ 1973, \apj, 186, 1007 
\bibitem{Zor10} Zorotovic, M., Schreiber, M.~R., G{\"a}nsicke, B.~T., \& Nebot G{\'o}mez-Mor{\'a}n, A.\ 2010, \aap, 520, A86 
\bibitem{Zor14}  Zorotovic, M., Schreiber, M.~R., Garc{\'{\i}}a-Berro, E., et al.\ 2014, \aap, 568, A68 
\end{thebibliography}
\end{document}